\documentclass[aps,prl,superscriptaddress,twocolumn]{revtex4-1}

\usepackage{bm} 
\usepackage{graphicx} 
\usepackage{amsmath}
\usepackage{amsthm,stmaryrd,mathtools}
\usepackage{amssymb} 
\usepackage{epstopdf} 
\usepackage{txfonts}
\usepackage{graphicx}
\usepackage{siunitx}
\usepackage{hyperref}
\usepackage{braket}
\usepackage{amsmath}

\usepackage{graphicx,graphics,times,color}
\usepackage{bm}
\usepackage{longtable}
\usepackage{color}
\usepackage{lipsum}
\usepackage{graphicx}
\usepackage{epstopdf}

\def\be{\begin{equation}}
\def\ee{\end{equation}}
\def\ba{\begin{eqnarray}}
\def\ea{\end{eqnarray}}

\begin{document}

\title{Measurement quench in many-body systems}

\author{Abolfazl Bayat}
\affiliation{Institute of Fundamental and Frontier Sciences, University of Electronic Science and Technology of China, Chengdu 610051, China}
\affiliation{Department of Physics and Astronomy, University College London, London WC1E 6BT, United Kingdom}

\author{Bedoor Alkurtass}
\affiliation{Department of Physics and Astronomy, University College London, London WC1E 6BT, United Kingdom}
\affiliation{Department of Physics and Astronomy, King Saud University, Riyadh 11451, Saudi Arabia}

\author{Pasquale Sodano}
\affiliation{International Institute of Physics, Universidade Federal do Rio Grande do Norte, 59078-400 Natal-RN, Brazil}

\author{Henrik Johannesson}
\affiliation{Department of Physics, University of Gothenburg, SE 412 96 Gothenburg, Sweden}
\affiliation{Beijing Computational Science Research Center, Beijing 100094, China}

\author{Sougato Bose}
\affiliation{Department of Physics and Astronomy, University College London, London WC1E 6BT, United Kingdom}

\date{\today}

\begin{abstract}
Measurement is one of the key concepts which discriminates classical and quantum physics. Unlike classical
systems, a measurement on a quantum system typically alters
it drastically as a result of wave function collapse. Here we suggest that this feature can be exploited for inducing
quench dynamics in a many-body system while 
leaving its Hamiltonian
unchanged. 
Importantly, by doing away with dedicated macroscopic devices for inducing a quench -- using
instead the indispensable measurement apparatus only -- the protocol is expected to be easier to implement and more resilient against decoherence.
By way of various case studies, we show that our scheme also has decisive advantages beyond reducing
decoherence -- for spectroscopy purposes and probing nonequilibrium scaling of critical and quantum impurity many-body systems.  
\end{abstract}


\maketitle

\emph{Introduction.--} 
Measurement is a fundamental concept which discriminates between the classical and quantum worlds. While in the classical regime measurement is noninvasive with no effect on the system, in the quantum domain, however, acquiring information, even through a local measurement, comes at the cost of an abrupt wave-function collapse which affects the entire system. 
The fundamental tests for the validity of quantum mechanics, such as the violation of Bell~\cite{bell1964einstein,aspect1981experimental} and Legett-Garg~\cite{leggett1985quantum} inequalities are based on quantum measurements. Moreover, they are crucial ingredients of almost all emerging quantum technologies such as quantum teleportation~\cite{bennett1993teleporting}, measurement-based quantum computation~\cite{briegel2009measurement}, fault-tolerant quantum computation \cite{lo1998introduction} and spin-chain quantum communication \cite{shizume2007quantum,pouyandeh2014measurement,bayat2015measurement}.  

All quantum protocols consist of preparation, manipulation and readout of one or more particles. While preparation and manipulation can be achieved by different means, the readout is unequivocally accomplished by measurements. Experimentally, these all rely on macroscopic devices which may induce decoherence and increase the complexity of the process. This raises the question: Is it possible to simplify the whole process by keeping only the indispensable part of the macroscopic devices, i.e. the measurement apparatus, for the complete preparation and manipulation of the system?

To answer the question, one should first recall that a key task of any quantum protocol is to induce the ``right'' kind of dynamics on a system. A particularly important  class is that of quench dynamics, where the time evolution is induced in the system by a sudden change in the Hamiltonian. Quantum quench physics has been the subject
of  extensive studies~\cite{polkovnikov2011colloquium,eisert2015quantum}, addressing fundamental problems such as equilibration~\cite{rigol2007thermalization,gogolin2016equilibration} and emergence of highly entangled  states~\cite{barmettler2009relaxation} to practical applications such as creating long-distance entanglement~\cite{bayat2010entanglement}. 
Experimentally, various features of quench dynamics have been observed in optical lattices~\cite{cheneau2012light,zeiher2017coherent,fukuhara2013quantum,fukuhara2013microscopic}, optical tweezers~\cite{bernien2017probing}, ion-traps~\cite{jurcevic2014observation,richerme2014non},  nuclear magnetic resonance devices~\cite{rao2014efficient} and coupled optical
fibers~\cite{bellec2012faithful,perez2013coherent}.

\begin{figure} \centering
	\includegraphics[width=6cm,height=3.5cm,angle=0]{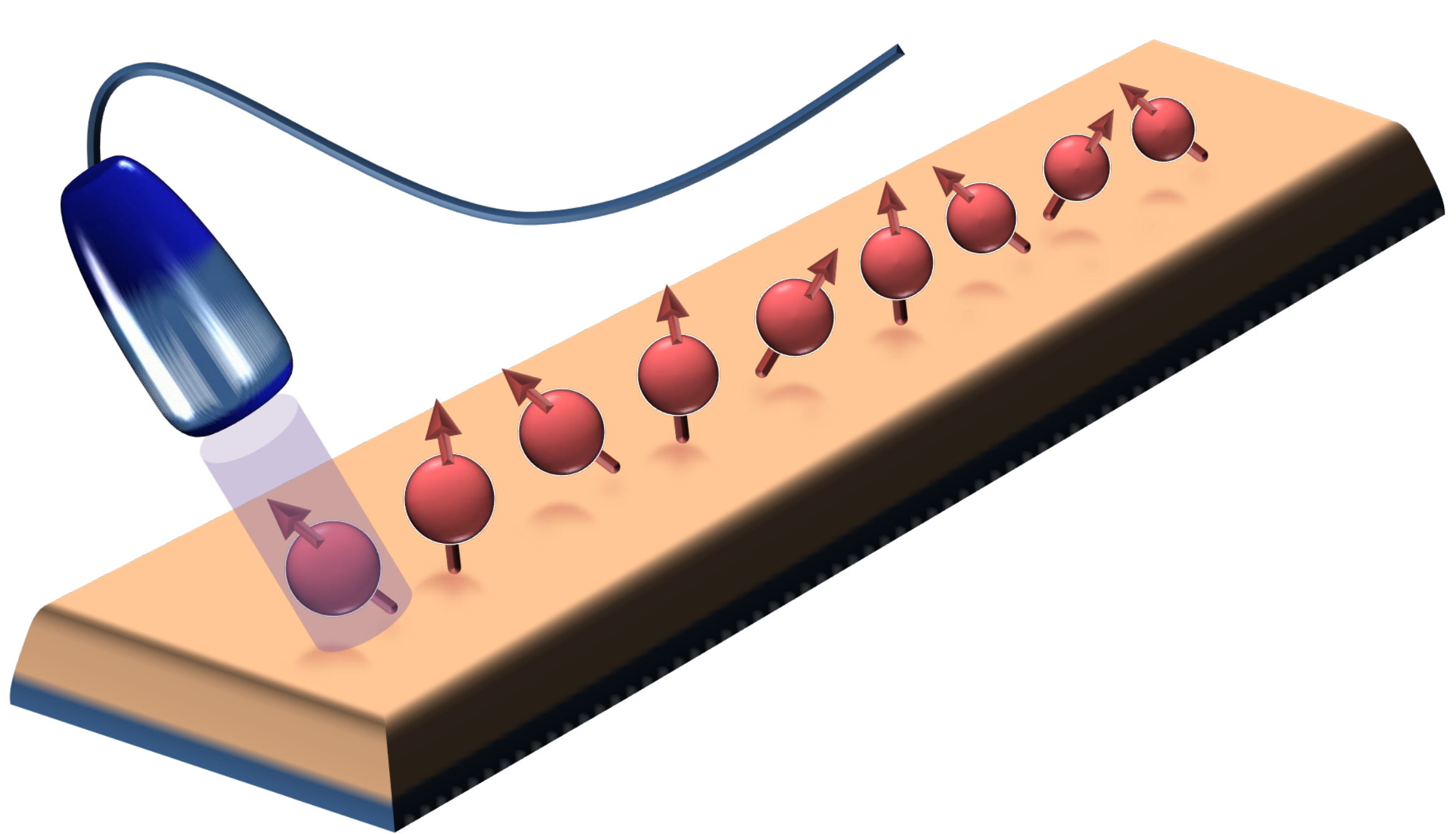}
	\caption{ \textbf{Measurement quench.} A local measurement is performed on one of the qubits of a chain prepared in its ground state. This causes a collapse to a new state and induces dynamics in the system.}
	\label{fig1}
\end{figure}

Here, we show that nonequilibrium  dynamics can be induced by a
local measurement, a {\em measurement quench}, keeping the Hamiltonian intact. We introduce several applications for this effect. Experimentally, a measurement
quench uses the same measurement device that is used for
reading the output signal, thus dispensing
with the need for dedicated devices for
inducing dynamics and therefore reduces decoherence. 
Note that, due to the wave function collapse, after a measurement quench the reduced density matrix of  {\em all} subsystems change abruptly. This is very different from creating excitations via local rotations~\cite{jurcevic2014observation,fukuhara2013quantum,fukuhara2013microscopic} in which only the local reduced density matrix of the rotated particles changes. Nonetheless, the same technology that performs local rotations can perform local measurements as well.

\emph{Measurement quench.--}
The notion of a measurement
quench is most easily introduced by way of example. For
this purpose, let us consider a chain of $N$ qubits interacting
through a many-body Hamiltonian $H$. The system is initialized in its ground state $|E_0\rangle$. We then measure the magnetization of one of
the qubits, say qubit $j$, in a certain basis which we here take as the Pauli $\sigma^x$ (cf. Fig.~\ref{fig1}). The measurement is encoded by the two projectors 
\begin{equation} \label{Projectors}
  \Pi^\uparrow_j{=}|\uparrow_j\rangle \langle \uparrow_j |\otimes I_{rest}, \quad \quad \Pi^\downarrow_j{=}|\downarrow_j\rangle \langle \downarrow_j |\otimes I_{rest},
\end{equation}
where $\uparrow_j$ and $\downarrow_j$ represent the outcomes of the measurement in the $x$-direction at site $j$ and $I_{rest}$ denotes the identity operator in the space of all the other qubits. According to the outcome of the measurement, at time $t{=}0$, the wave function of the full system collapses to one of the following quantum states:
\begin{equation} \label{psi_0}
	|\Psi^\mu_j(0)\rangle{=}\Pi^\mu_j |E_0\rangle/(p_j^\mu)^{1/2},  \quad p^\mu_j{=} \langle E_0| \Pi^\mu_j |E_0\rangle, \quad \mu{=}\uparrow,\downarrow.
\end{equation}
Here $p^\mu_j$ is the probability of having the outcome $\mu_j$ for the measurement. Since the new quantum state is no longer an eigenstate of the Hamiltonian the system starts to evolve as 
\begin{equation} \label{psi_t}
|\Psi^{\,\mu}_j(t)\rangle{=}e^{-iHt}|\Psi^{\,\mu}_j(0)\rangle=\sum_n e^{-iE_nt}|E_n\rangle \langle E_n|\Psi^{\,\mu}_j(0)\rangle,
\end{equation} 
where $E_n$ and $|E_n\rangle$ (for $n{=}0,1,\cdots$) are the eigenvalues and the eigenstates of $H$, respectively. 
Without loss
of generality, from now on we assume that the outcome of the measurement is $\mu=\,\uparrow$ and drop the symbol $\mu$. The magnetization of the measured qubit $j$ at any later time $t$ is then given by $m_j^x(t){=} \langle \Psi^\uparrow_j(t)| \sigma^x_j |\Psi^\uparrow_j(t)\rangle$. It follows from Eq.~(\ref{psi_t}) that  
\begin{equation} \label{mx_t}
m_j^x(t)=\sum_{n,m} e^{-i(E_n-E_m)t} \langle E_m|\sigma_j^x|E_n\rangle \langle E_n|\Psi_j^\uparrow(0)\rangle \langle \Psi_j^\uparrow(0)|E_m\rangle.
\end{equation}
To read  $m_j^x(t)$ one has to measure qubit $j$ again, which is the very same process that was used to induce the dynamics. 


\begin{figure} \centering
	\includegraphics[width=0.45\textwidth]{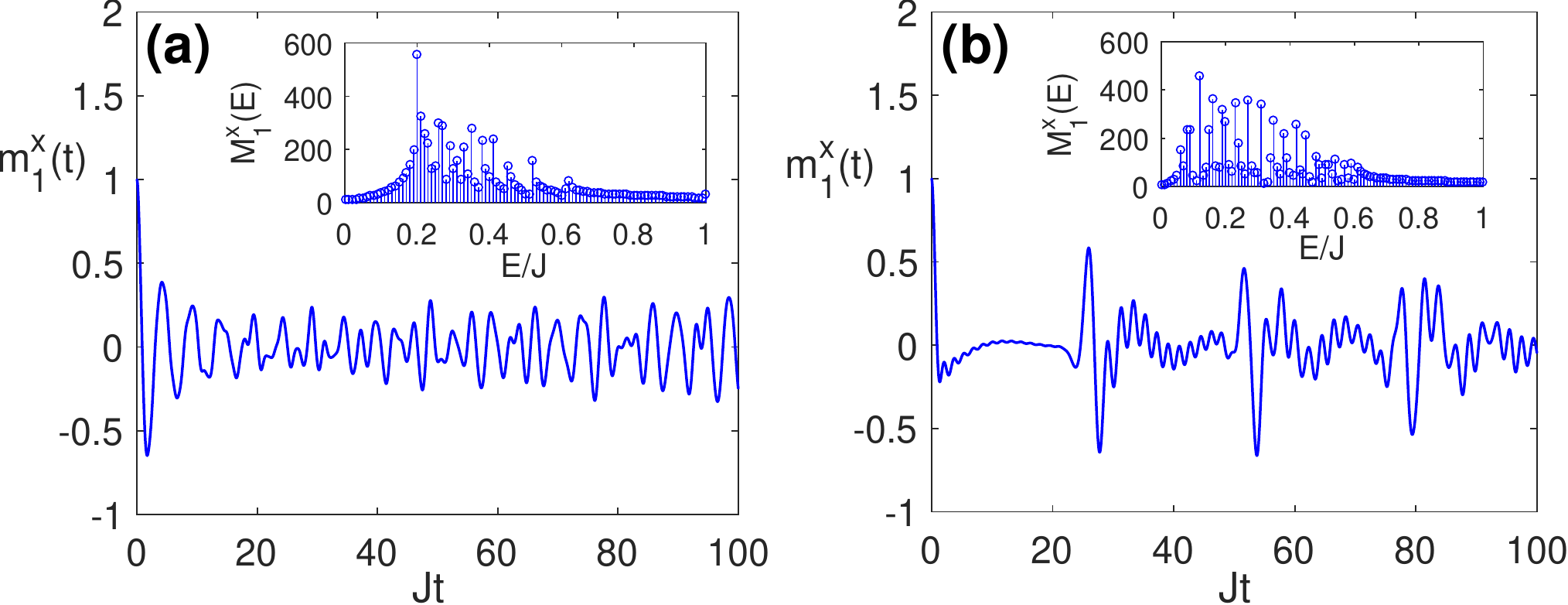}
	\caption{ \textbf{Transverse field Ising chain with long-range interaction.} The magnetization $m_1^x(t)$ as a function of time in a chain of length $N=20$ and $ B/J=1$ for two types of long-range interaction: (a) $\alpha=0.5$; (b) $\alpha=3$.  The insets show the Fourier transform $M_1^x(E)$ as function of energy $E$ for the chosen value of $\alpha$.}
	\label{fig_Ising_long}
\end{figure}

\emph{Application 1: Spectroscopy.--}
Quantum simulation~\cite{lloyd1996universal,cirac2012goals,blatt2012quantum} is one of the most important goals of quantum technologies. Recently, simulating many-body systems with more than 50 particles have been possible with both cold atoms~\cite{bernien2017probing} and trapped ions~\cite{zhang2017observation}. In fact, a wide range of spin Hamiltonians including the long-range Ising model~\cite{porras2004effective,friedenauer2008simulating} can be simulated in ion traps using either optical dipole forces~\cite{kielpinski2002architecture} or inhomogeneous magnetic fields~\cite{mintert2001ion,lekitsch2017blueprint,zippilli2014adiabatic}. Local addressability is also available in these ion trap technologies~\cite{naegerl1999laser,lekitsch2017blueprint}. Spurred by these advances, the dynamics of the long-range transverse field Ising  chain has attracted huge interest in both theory \cite{schachenmayer2013entanglement} and experiment~\cite{zhang2017observation}. The model is defined by~\cite{kim2009entanglement,richerme2013quantum}
\begin{align} \label{Ising_Long_range}
H=J \sum_{i\neq j}\frac{1}{|i-j|^\alpha} \sigma_i^x\sigma_j^x +  B \sum_i \sigma_i^z,
\end{align}
where $\sigma_i^{x}$ and $\sigma_i^{z}$ are the Pauli matrices acting on site $i$, $B$ is a magnetic field strength, and $\alpha$ determines the range of the interaction such that $\alpha=0$ makes the system fully connected while the limit $\alpha \rightarrow \infty$ represents the nearest-neighbor chain. The exchange coupling is here taken to be antiferromagnetic, $J>0$. Current experimental techniques allow $\alpha$ to
be tuned within the interval $0 \leq \alpha \leq 3$~\cite{monroe2015quantum}. Except for the special cases of $\alpha=0$~\cite{lipkin1965validity,russomanno2015thermalization} and $\alpha \rightarrow \infty$~\cite{sachdev2011quantum}, the Hamiltonian in (\ref{Ising_Long_range}) is not solvable and thus the spectrum can only be found for short chains through exact diagonalization. As shown in Ref.~\cite{schachenmayer2013entanglement}, starting from a product state, the entanglement entropy grows linearly for short-range interactions (i.e. large $\alpha$) and logarithmically for long-range interactions (small $\alpha$)~\cite{schachenmayer2013entanglement}.

As the Hamiltonian (\ref{Ising_Long_range}) commutes with the parity operator $P=\prod_{i=1}^{N} \sigma_i^{z}$, the ground state has always a definite parity. In particular, for even $N$, the ground state has even parity in which measuring $\sigma_j^x$ results in either $\uparrow_j$ or $\downarrow_j$ with equal probability. Eq.~(\ref{mx_t}) can then be simplified to 
\begin{equation} \label{Mzt_Uniform}
m_j^{x}(t)=\sum_{n} \cos\left[\left(E_n-E_0\right) t\right] \left| \langle E_n| \sigma_j^x| E_0\rangle \right|^2,
\end{equation}
in which the frequencies of the oscillations are determined only by the energy gaps between the
ground state $|E_0\rangle$ and those excited states $| E_n\rangle$ for which $|\langle E_n| \sigma_j^x| E_0\rangle|^2$ is nonzero. In Figs.~\ref{fig_Ising_long}(a)-(b) the local magnetization $m_1^{x}(t)$ is plotted versus time for $\alpha=0.5$ and $\alpha=3$ respectively in a system of length $N=20$, using open boundary conditions. 
As the figures show, for the larger $\alpha$ the dynamics equilibrates after a short oscillation and then revives due to the finite size of the system, while for the smaller $\alpha$ there
is no hint of equilibration due to more frustration in the staggered ordering of the spins.

One may compute  the Fourier transform of the local magnetization as $M_j^x(E)=\frac{1}{2\pi} \int_0^\infty m_j^x(t) e^{-iEt}  dt$, which takes the form  

\begin{equation} \label{Mzt_Uniform}\nonumber
M_j^{x}(E)=\sum_{n} \left| \langle E_n| \sigma_j^x| E_0\rangle \right|^2 \bigl[ \delta(E-E_n+E_0)+\delta(E-E_0+E_n) \bigr]/2. 
\end{equation}
In the insets of Figs.~\ref{fig_Ising_long}(a)-(b) we have plotted $M_1^x(E)$ for their respective dynamics. As is evident from the figures, more frequencies are excited for larger $\alpha$, resulting in the equilibration of $m_1^x(t)$ already on short time scales. The location of the peaks of $M_1^{x}(E)$ correspond to the energy gaps $E_n{-}E_0$ and thus $M_1^{x}(E)$ can be used as an efficient  spectroscopic probe of an \emph{unsolvable} system. In order to capture more eigenvalues of the Hamiltonian one can perform the measurement quench on other sites $j\!\neq \!1$ by
which other eigenstates also get excited.

\begin{figure} \centering
	\includegraphics[width=0.45\textwidth]{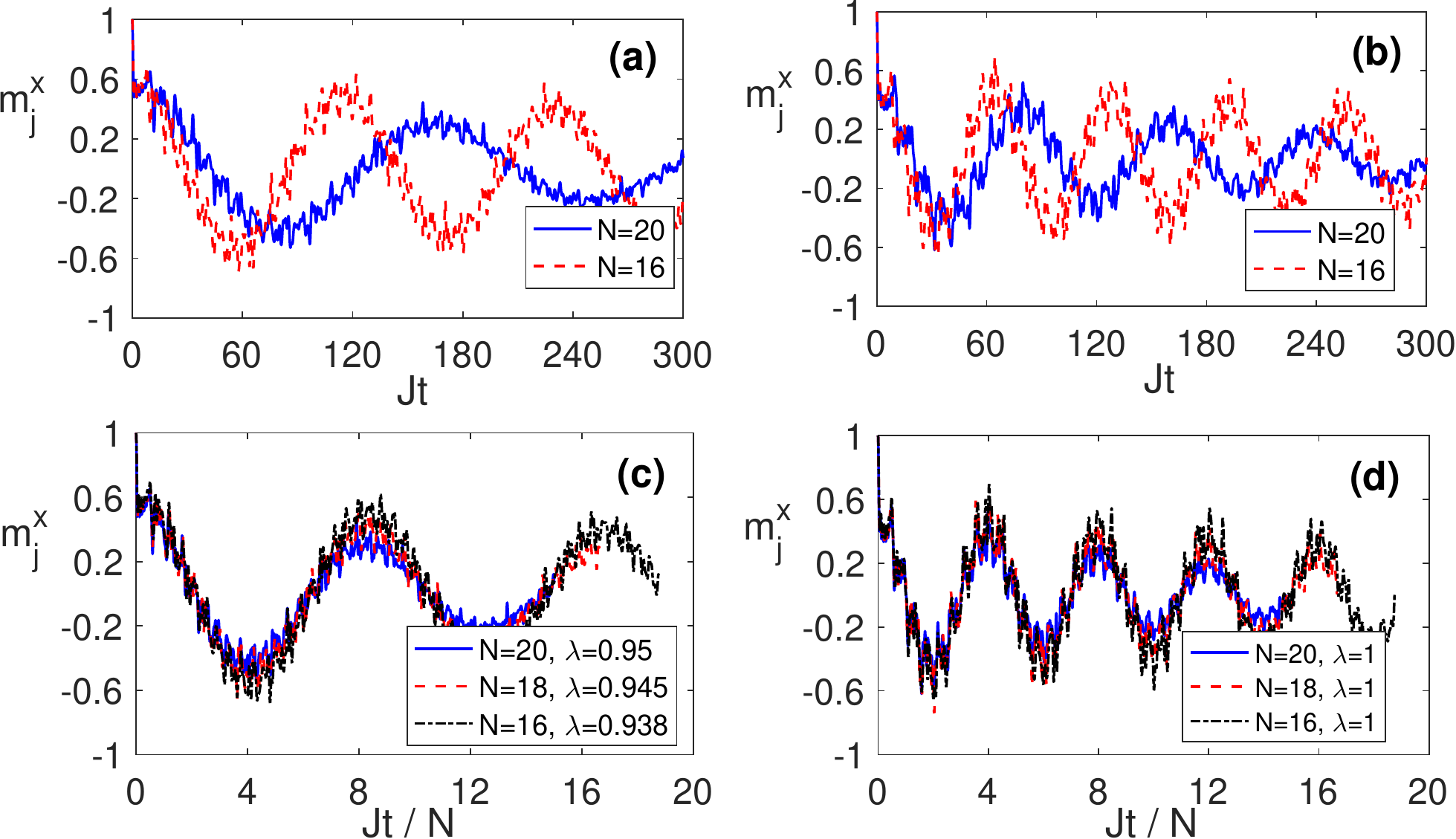}
	\caption{ \textbf{Magnetization dynamics in the transverse field Ising chain.} Time evolution of the local magnetization after a measurement quench in the Ising model for: (a) $\lambda\!=\!0.95$; and (b) $\lambda\!=\!\lambda_c=\!1$. Nonequilibrium scaling of the local magnetization $m_j^x$ versus $t/N$ for (c) $|\lambda-\lambda_c|\simeq 1/N$ and (d) $|\lambda-\lambda_c|=0$, for which all lengths behave in the same way since $\xi\sim N$.
}
\label{fig3}
\end{figure}

\emph{Application 2: Nonequilibrium scaling near a quantum phase transition.--}
Scaling is a key feature in many-body systems near criticality~\cite{sachdev2011quantum}. 
Maybe the best known example from out-of-eqilibrium dynamics is the Kibble-Zurek scaling~\cite{kibble1976topology,zurek1985cosmological,kibble2007phase}, present when a many-body system is driven through a continuous phase transition at a finite rate. 
In a like manner, the asymptotic approach to equilibrium of a system that is suddenly quenched close to a quantum critical point is expected to be governed by equilibrium critical exponents \cite{DeGrandi2010}. 
Scaling in a critical system that has been subject to a global quench is expected only after long times~\cite{kennes2014universal}.
Local Hamiltonian quenches \cite{Eisler2007,Calabrese2007} also exhibit power-law scaling with time and distance from the quench \cite{Calabrese2007,Stephan2011}. Interestingly, the corresponding static scaling dimensions can be extracted at intermediate time scales \cite{Divakaran2011,Igloi2014}.  With this as a backdrop, one may inquire how a local measurement quench gives rise to scaling of observables. 

To find out, let us consider the transverse-field Ising chain (TFIC). This model, solvable via Jordan-Wigner transformation, serves as a paradigm for quantum phase transitions~\cite{sachdev2011quantum}. The Hamiltonian, masked as the limit $\alpha \!\rightarrow\!\infty$ in Eq.~(\ref{Ising_Long_range}), 
is 
\begin{equation} \label{Hamiltonian_ising}
H=\sum_{i=1}^N \sigma^x_i \sigma^x_{i+1} + \lambda  \sum_{i=1}^N\sigma^z_i\, ,
\end{equation}
where the dimensionless parameter $\lambda$ plays the role of $B/J$ in the Hamiltonian (\ref{Ising_Long_range}), and here, in contrast to the previous example, periodic boundary conditions are imposed.  By varying $\lambda$, the system undergoes a quantum phase transition at $\lambda\!=\!\lambda_c\!=\!1$, from an antiferromagnetic ($\lambda{<}\lambda_c$) to a paramagnetic ($\lambda{>}\lambda_c$) phase. Near the critical point the correlation length diverges as $\xi {\sim} |\lambda-\lambda_c|^{-\nu}$, with the exponent $\nu{=}1$~\cite{sachdev2011quantum}.   

By translational invariance it does not matter on which site $j$ the measurement quench is performed. The well-defined parity of the TFIC eigenstates implies that $m_j^x(t)$ follows the Eq.~(\ref{psi_0}).
Unfortunately, the correlation functions $| \langle E_n| \sigma_j^x| E_0\rangle |^2$ in Eq.~(\ref{psi_0}) 
cannot be expressed in terms of a finite number of
free fermionic correlation functions in a periodic chain~\cite{barouch1971statistical}. This makes it
difficult to benefit from a Jordan-Wigner transformation and instead we resort to numerical exact diagonalization. 
The result for $m_j^x(t)$ is plotted in Figs.~\ref{fig3}(a) and (b) for $\lambda=0.95$ and $\lambda=\lambda_c=1$, respectively. The magnetization exhibits persistent small high-frequency fluctuations on top of a global oscillating low-frequency signal which decays slowly with time, suggesting gradual equilibration. 

It is worth mentioning that the local magnetization $m_j^x$ is a non-equilibrium quantity and does not serve as an order parameter for the system. Therefore, it is not clear whether one can see scaling behavior for $m_j^x$. In fact, one may consider $m_j^x$ as a function of $t$, $N$ and $\xi$ where the dependence on $\lambda$ has been replaced by $\xi$ using $\xi \sim |\lambda-\lambda_c|^{-\nu}$. 
Scaling in the time evolution of $m_j^x$ means that it is not a function of $t, N$ and $\xi$ independently, but instead is parameterized as $m_j^x(t/N,N/\xi)$.
To verify this, one fixes $N/\xi$ and then plots $m_j^x$ as a function of $t/N$ for various system sizes so that all curves collapse on top of each other. In order to fix $N/\xi$ one can choose $\lambda$ for each system size $N$ such that $N|\lambda-\lambda_c|^\nu$ remains fixed. In Fig.~\ref{fig3}(c) such a data collapse is shown for three different system sizes. As the figure shows, that happens when $\nu$ is chosen to be $\nu=1$, i.e. the critical exponent known for the TFIC. 
An interesting case is $\lambda=\lambda_c$ for which all system sizes collapse on each other as in this case $N|\lambda-\lambda_c|^\nu$ becomes zero.
As evident from Figs.~\ref{fig3}(c) and (d), while the low-frequency signals show perfect data collapse, this is not so for the high-frequency fluctuations.
This suggests that the two frequency components of $m_j^x$ have distinct characteristics: (i) the low-frequency part shows universal scaling behavior and thus exhibits perfect data collapse; and (ii) the small high-frequency part is non-universal and does not scale.
Interestingly, the measurement quench is very different from the local Hamiltonian quenches at a defect in a critical TFIC in which the exponents are nonuniversal and  vary with the defect parameter~\cite{Igloi2014}.

\emph{Application 3: Nonequilibrium scaling in the Kondo model.--}
The Kondo model \cite{Andrei1983,hewson1997kondo,Pustilnik2004} serves as a paradigm for electronic many-body systems where the interaction with a quantum impurity dynamically generates a length scale. Taking advantage of the presence of this scale -- the Kondo screening length $\xi_K$ -- allows data from different systems to be collapsed onto a single curve, similar to a critical system like the TFIC discussed above. Theoretical work \cite{Ralph1994,Costi1994,Schiller1995}, as well as transport measurements on quantum dots in the Kondo regime \cite{Goldhaber2008}, show that universal scaling behavior is maintained in nonequilibrium, and a Kondo cloud can form as the result of a local quench
between an impurity and  an electron gas~\cite{Nuss2015}.  
Moreover, a local Hamiltonian quench in a Kondo system may provide distance-independent entanglement between two distant impurities~\cite{bayat2010entanglement}. 
Here we add to the nonequilibrium picture of the Kondo physics, using a measurement quench.

\begin{figure} \centering
	\includegraphics[width=8cm,height=4cm,angle=0]{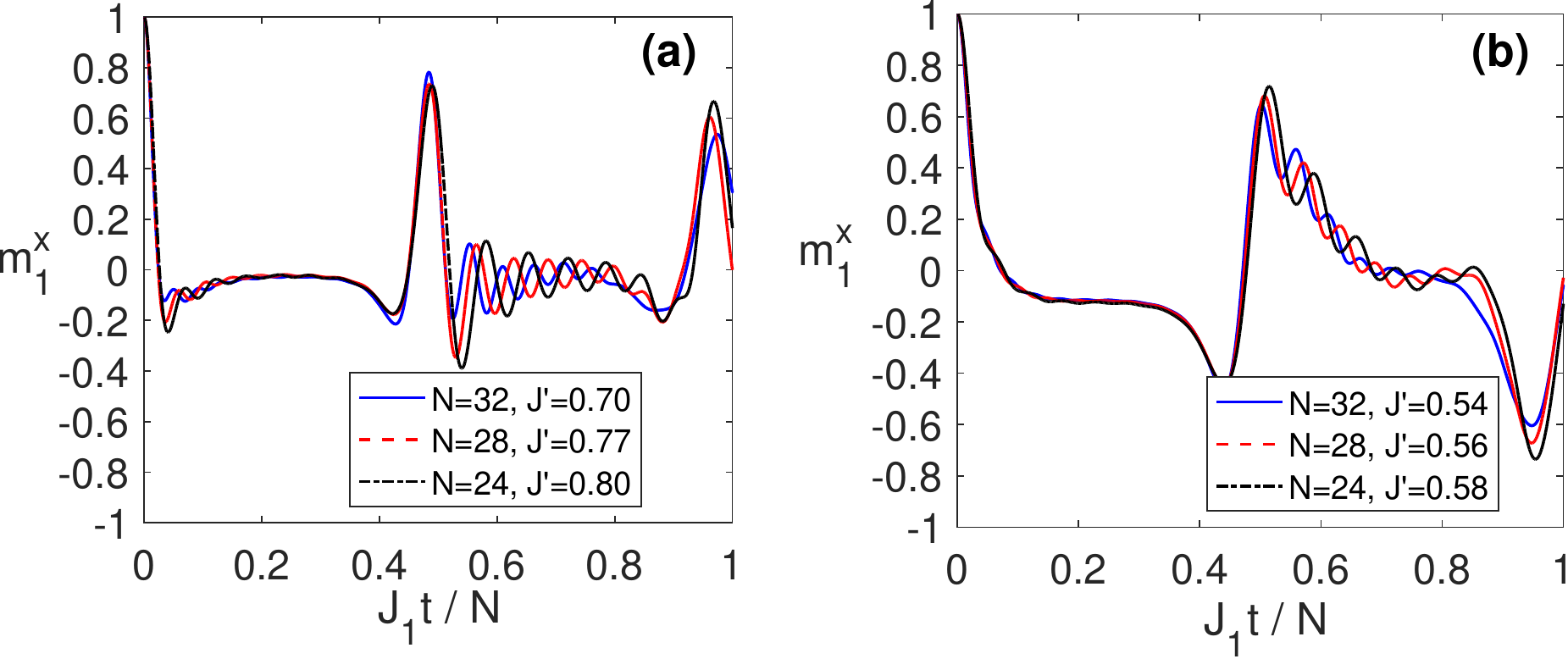}
	\caption{ \textbf{Magnetization dynamics in the Kondo model.} Nonequilibrium scaling of the local magnetization $m_1^x$ versus $t/N$ when the length scale $\xi_K$ is tuned to: (a) $N/\xi_K\simeq 3.4$; (b) $N/\xi_K\simeq 2$.}
	\label{fig4}
\end{figure}

For this purpose, it is more convenient to use a spin-chain emulation \cite{laflorencie2008kondo} of the Kondo model, allowing for efficient computations~\cite{bayat2010negativity} using the Density Matrix Renormalization Group (DMRG)~\cite{white1992density,white1993density}. The spin chain has the Hamiltonian 
\begin{equation} \label{Hamiltonian-SIKM}
H=J'(J_1 \boldsymbol{\sigma}_1\cdot\boldsymbol{\sigma}_2+J_2\boldsymbol{\sigma}_1\cdot\boldsymbol{\sigma}_3)+
J_1\sum_{i=2}^{N-1} \boldsymbol{\sigma}_i\cdot\boldsymbol{\sigma}_{i+1}+J_2\sum_{i=2}^{N-2} \boldsymbol{\sigma}_i\cdot\boldsymbol{\sigma}_{i+2}
\end{equation} 
where $J_1$ ($J_2$) is the (next-) nearest neighbor coupling and the dimensionless parameter $J'$ represents the impurity coupling, with the impurity located at site $i=1$. By fine tuning $J_2/J_1 = 0.2412$ to the critical point of the spin-chain dimerization transition \cite{laflorencie2008kondo}, the Hamiltonian in (\ref{Hamiltonian-SIKM}) provides a faithful representation of the spin sector of the Kondo model \cite{laflorencie2008kondo}. 
The Kondo screening length $\xi_K$ can be identified with the spatial extent of a block of spins with which the impurity is maximally entangled \cite{bayat2010negativity}. Assuming that the number of sites on the chain is even, the SU(2) symmetry of the model implies that $p_j^\uparrow=1/2$.  A measurement quench is now performed on the impurity spin at site $j=1$. To compute the subsequent time evolution, we employ exact diagonalization for short chains up to $N=20$ and time-dependent Runge-Kutta DMRG simulation for longer chains~\cite{feiguin2005time}. Following the same finite-size scaling procedure as we used for the TFIC, we obtain the results displayed in Figs.~\ref{fig4}(a)-(b), with $m_1^x(t)$ plotted versus $t/N$ when $N/\xi_K$ is fixed to $N/\xi_K=3.4$ and $N/\xi_K=2$, respectively. The value of $\xi_K$ is found using the entanglement approach of Ref.~\cite{bayat2010negativity}.
%
%
As one can see, there are two distinct regions: (i) a scaling region which starts from $t=0$ and extends to $J_1t\backsimeq N/2$ over which there is an almost perfect data collapse; and (ii) a finite-size region $J_1t>N/2$ over which the data collapse gets distorted due to reflection of excitations from the boundaries.  Note that the different scaling behaviors seen in the Kondo and TFIC models is due to the fact that the length scale $\xi_K$ is dynamically generated, and its divergence for small $J'$ does not reflect a quantum phase transition.

It is worth pointing out that despite the success of DMRG to capture the low-energy sector of a many-body system, such as the Kondo model, it cannot compete with a real quantum simulator. First, as entanglement grows, the DMRG algortihm fails to give an accurate description~\cite{trotzky2012probing}. Secondly, even for models where DMRG performs at its best the time scale to compute the evolution is order of magnitudes larger compared to monitoring the same dynamics on a real many-body system, like a quantum simulator using cold atom, trapped ions, or photonics. For such real-time experiments our measurement quench protocol is expected to come into its own.
		

\begin{figure} \centering
	\includegraphics[width=8cm,height=4cm,angle=0]{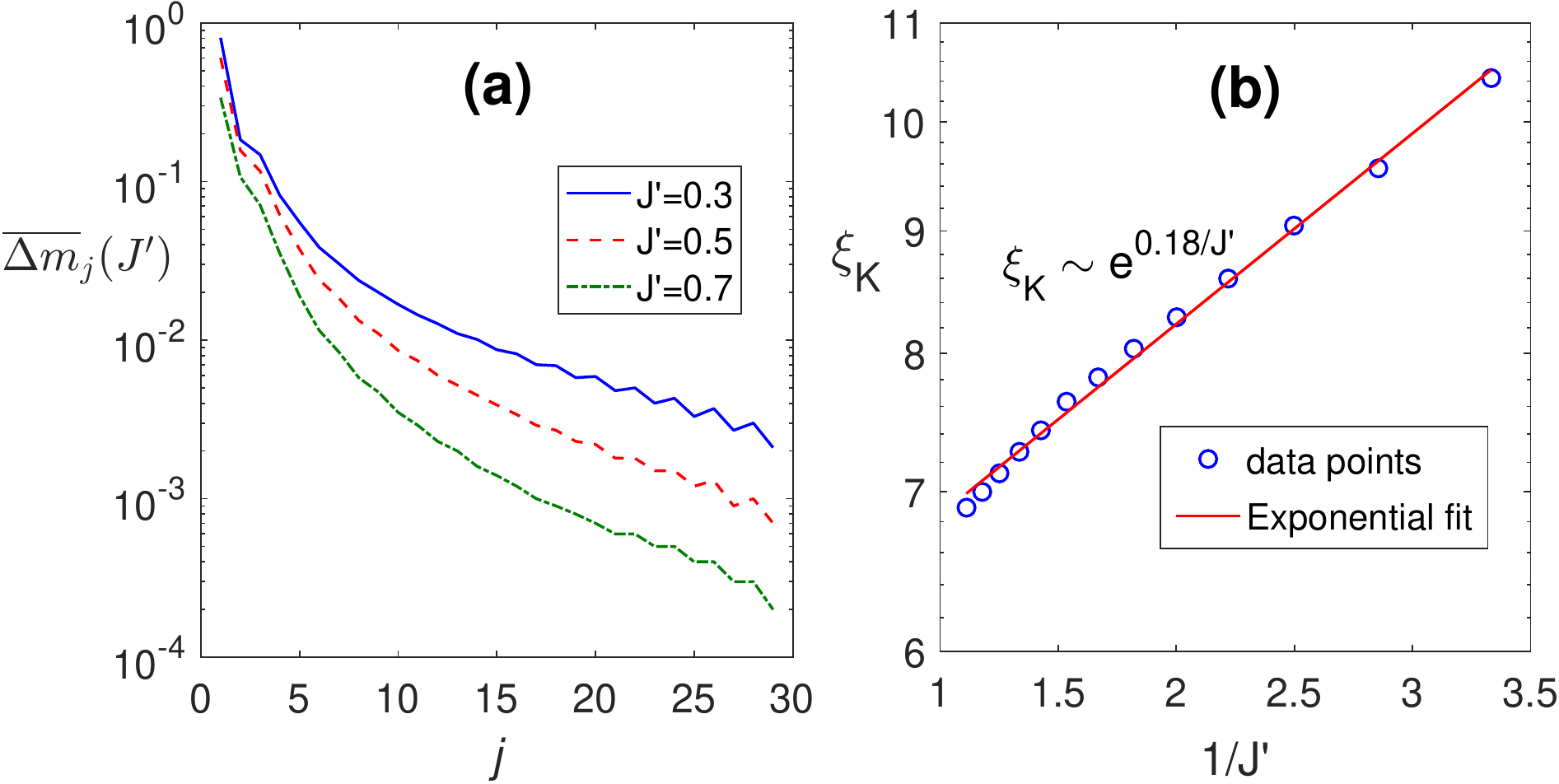}
	\caption{ \textbf{Detecting the Kondo cloud.} (a) Average local magnetization difference $\overline{\Delta m}_j$ as function of measurement site $j$ in a chain of $N=32$ and $J'=0.3$. 
	(b) The length scale $\xi_K$ versus $1/J'$ for the chain of length $N=32$ and $J'=0.3$. The semilogarithmic plot confirms its exponential dependence with the fitting parameter $A=0.18$.}
	\label{fig5}
\end{figure}

\emph{Application 4: Detecting the Kondo screening cloud.--} 
While much is known about Kondo physics \cite{Andrei1983,hewson1997kondo,Pustilnik2004}, the experimental detection of the Kondo screening cloud, of size $\xi_K$, remains a challenge \cite{Affleck2010}. We here suggest a new type of protocol for determining $\xi_K$, based on a measurement quench. To do so, we consider the Hamiltonian in Eq.~(\ref{Hamiltonian-SIKM}) and perform a measurement quench on the spin at site $j$.
If this site is far from the impurity, outside the Kondo cloud, then the time evolution of the magnetization $m_j^x(t)$ is not affected by the presence of the impurity on short time scales. Thus, by comparing the evolution of $m_j^x(t)$ in the presence $(J^{\prime}\!\neq\!1)$ and absence $(J^{\prime}\!=\!1)$ of the impurity, one expects that there is no difference between the two cases when the site resides outside the Kondo cloud. Guided by this, we define an average local magnetization difference as 
\begin{equation} \label{ma_av_j}
\overline{\Delta m}_j(J')=\frac{1}{T}\int_0^T \left| m_j^x(J')-m_j^x(J'\!=\!1) \right| dt,
\end{equation}
where $T=1/J'$ is a short time compared to the time needed for spin excitations to propagate across the chain. 
In Fig.~\ref{fig5}(a) we plot $\overline{\Delta m}_j(J')$ as a function of $j$ for different values of $J'$ in a spin chain of length $N=32$. As one can see from the figure, $\overline{\Delta m}_j(J')$ decays exponentially when the site $j$ is far away from the impurity site $j\!=\!1$. By considering an exponential fitting function of the form $\overline{\Delta m}_j(J') {\sim} e^{-j/\xi_K}$ 	to the tail of the data 
one can extract the length scale $\xi_K$. In Fig.~\ref{fig5}(b) we plot $\xi_K$ thus obtained as a function of $1/J'$. Choosing $A=0.18$, one obtains very good agreement 
with $\xi_K {\sim} e^{A/J'}$, the expected exponential scaling for the Kondo screening length \cite{hewson1997kondo}.

\emph{Conclusion.--} We have shown that a local measurement can be harnessed to induce nonequilibrium dynamics in many-body systems. 
In contrast to conventional quench protocols, where the Hamiltonian is manipulated, our proposal is easier to implement and less prone to decoherence. Several applications of measurement quenches have been discussed.  They allow for efficient spectroscopy of nontrivial spin systems, for extracting nonequilibrium scaling at quantum criticality and in quantum impurity systems, and also for probing the elusive screening cloud in the Kondo model. 
Various physical setups, in which projective measurements have been realized, can potentially implement our protocol, including ion traps~\cite{mintert2001ion,zhang2017observation,lekitsch2017blueprint}, optical lattices~\cite{sherson2010single,weitenberg2011single,fukuhara2012quantum,fukuhara2013microscopic}, Rydberg atoms~\cite{bernien2017probing} and superconducting circuits~\cite{barends2016digitized,roushan2017spectroscopic}. Importantly, a measurement quench can be performed on any initial state, including a thermal state and all the introduced applications remain valid if the temperature is small enough so that the measurement quench only excites low-energy eigenstates.

\emph{Acknowledgments.--} 
AB thanks the University of Electronic Science and Technology of China for their support. SB and AB acknowledge the EPSRC grant $EP/K004077/1$ and the ERC under Starting Grant 308253 PACOMANEDIA. 
HJ acknowledges support from the Swedish Research Council through Grant No. 621-2014-5972.


%

\end{document}